\documentclass{ws-procs9x6}
\usepackage{ws-multind}
\makeindex{author} 
\begin{document}

\wstoc{For proceedings editors: Combining contributions using\\
WS-procs9x6 master document in \LaTeX2e}{F. Author}

\title{Probing the space time around a black hole\\ with X-ray variability}

\author{Tomaso M. Belloni}
\index{author}{Belloni, T.M.} % or \aindx{Author, F.}

\address{INAF - Osservatorio Astronomico di Brera,\\
Via E. Bianchi 46, I-23807 Merate, Italy \\E-mail: tomaso.belloni@inaf.it}

\begin{abstract}
In the past decades, the phenomenology of fast time variations of high-energy flux from black-hole binaries has increased, thanks to the availability of more and more sophisticated space observatories, and a complex picture has emerged. Recently, models have been developed to interpret the observed signals in terms of fundamental frequencies connected to General Relativity, which has opened a promising way to measure the prediction of GR in the strong-field regime. I review the current standpoint both from the observational and theoretical side and show that these systems are the most promising laboratories for testing GR and the observations available today suggest that the next observational facilities can lead to a breakthrough in the field.
\end{abstract}

\keywords{X-ray binaries; accretion; time variability; General Relativity}

\section{Introduction: the promise of X-ray binaries}

X-ray binaries are stellar binary systems that contain a normal star and a collapsed object, either a black hole or a neutron star, in which the strong gravitational pull of the latter strips matter from the companion and accretes it. The process of accretion is complex and not fully understood. Because of its angular momentum, the gas cannot fall directly onto/into the compact object, but forms a disk around it. Matter at each radius follows a Keplerian orbit, slowly spiralling towards the center due to friction, emitting radiation at higher and higher energies as it nears the collapsed object. The inner parts of the accretion flow, below $\sim$100 R$_g$, are very hot and strong X-ray emission is observed. The structure of the accretion flow is very complex and varies with time depending on the rate of mass flowing through it. Although the supply from the companion star is expected to be constant, instabilities in the accretion flow result in a variable accretion rate through the inner parts of the disk. This variability can be extreme in transient systems, in which the compact object spends most of its time accreting at a very low rate, to experience surges of accretion for periods of weeks to months when the observed high-energy luminosity increases by several orders of magnitude\cite{BelloniMotta16}. 

An important aspect of the study of X-ray binaries is the determination of the nature of the compact object. While direct evidence for the presence of a neutron star can be obtained from the observation of high-frequency pulsations or thermonuclear X-ray bursts (caused by unstable nuclear burning on the surface of the object, indication that a surface is present), the black-hole nature of the central object has not so far been ascertained with direct measurements\cite{TMB18}. The strongest indication for a black hole is indirect, namely the measurement of the mass of the compact object from studying the optical modulation from the binary system\cite{CasaresJonker}.

To determine the presence of a black hole directly, we need to be able to observe in the X-ray emission, which originates very close to the black hole, effects due to General Relativity (GR), which would also allow us to check the validity of the theory in the strong field regime. The best approach would be to measure the mass of the compact object directly, with additional ideal observables being GR precession frequencies of orbiting matter and the presence of an innermost stable circular orbit (ISCO). The absence of a solid surface would be very difficult to prove as it is a negative measurement.
For the identification and measurement of GR effects in the strong field, X-ray binaries are our best laboratories. Active Galactic Nuclei, which contain supermassive black holes, are not a match as their gravitational potential is equivalent, but the field curvature much less\cite{Psaltis}. Spectacular measurements are being provided by the observation of the double pulsar\cite{kramer}. However, the two neutron stars in the system, that can be treated as test particles, are 7$\times 10^5$km apart. The same applies to the original Hulse\&Taylor binary pulsar\cite{weisberg}. In the case of a black-hole binary, our test particles orbit the object at a few gravitational radii, but they are not really test particles, but plasma in a complex accretion flow whose properties are not completely known. 
Nevertheless, these laboratories are available thanks to X-ray astronomical satellites that provide high-energy data of increasing quality (and quantity) and are now giving us the first answer. In this chapter, I will concentrate on the possibilities offered by the analysis of time variability in the X-ray flux, which close to the black hole takes place on time scales well below a second.

\section{Spectral approaches}

Before introducing time variability, it is important to mention other methods that are being used to estimate GR parameters based on the analysis of emitted X-ray spectra\cite{MillerMiller}. These methods can be very powerful and are yielding more and more refined estimates of black-hole spins.

\subsection{Continuum spectra}

The energy spectra originating from black-hole binaries (hereafter BHB) are very complex, being the superposition of different components from different regions of the accretion flow, and time dependent. One component that is observed is produced in a geometrically-thin and optically-thick accretion disk, as modelled originally by Shakura \& Sunyaev\cite{SS}. The spectrum emitted by such a disk is the superposition of blackbody components from different radii, with a specific radial temperature distribution. The integrated spectrum has the luminosity of a sphere whose radius is the inner radius of the disk and whose temperature is that at the inner radius. Essentially, it behaves like a black body, which means that an estimate of its luminosity and temperature directly translates into a measurement of the inner radius of the disk. Given a black-hole mass (estimated from optical data), the radius can be expressed in gravitational radii and if it corresponds to the ISCO it yields the spin of the black hole. 
The emitted spectrum is more complex than a simple sum of blackbody components and more sophisticated and realistic models have been produced\cite{Davis}.

Despite the (variable) presence of several other spectral components, BHBs in their so-called ''high-soft state'' emit a spectrum that is consistent with an almost pure thermal component from the accretion disk. Detailed modelling showed that the estimated inner disk radius is constant and can therefore be associated to the ISCO. This has led to the measurement of the back-hole spin for a number of sources\cite{Jeff}. In order to measure the spin one must assume that the measured inner radius corresponds to ISCO, but any larger radius would lead to a higher spin. However, in order to measure the radius one needs to model the spectrum very accurately, take care of possible weak contamination from additional components, assume a distance to the source and assume an inclination of the system (the disk is not a sphere and its luminosity depends on its inclination, which of course is a random parameter and can be estimated with optical measurements). The presence of interstellar and local absorption also complicated the measurement.

\subsection{Iron line emission}

A different and very powerful method used to estimate black-hole spins is related to an additional component of the energy spectrum. The inner region of the accretion disk emits radiation that also irradiates on the disk. The incident X-ray radiation is ``reflected'' by the disk and emits a spectrum that is composed of fluorescence lines, of which the most prominent are Fe K lines, photoelectric absorption from the Fe shell above the lines, and an increase in form of a continuum above that, with a gradual flux decrease at high energies. This spectrum is complex, but it is dominated by an iron emission line that in a cloud of gas would be very narrow. However, an accretion disk is made of matter orbiting at the local Keplerian period, which means that the overall spectrum will be the superposition of spectra from different annuli. The spectrum from each annulus will be modified by Doppler effect due to the fast rotation, by relativistic redshift and boosting. The final shape is expected to be very broad and possibly double-horned\cite{RossFabian,FabianRoss}. Of course all these effects affecting the line depend on the inclination, but it turns out that the blue wing is sensitive mostly to the inclination and the red wing to how close the disk comes to the black hole because of the strong redshift. Therefore, detecting a broad line excess and modelling it with the proper model can potentially yield a measurement of ISCO. Several measurements have been obtained, in some cases combined with the continuum method described above\cite{Steiner}. This method must deal with the same issues as the one above, with the exception of the distance measurement, as radii are obtained directly in units of gravitational radii. In particular, the full energy spectrum extends over a broad X-ray band and is very complex, with several components some of which overlap in the energy region where the broad line is observed (6-7 keV). Since a very broad band extending from 3 keV to 7 keV is observationally a continuum component, this  means that the full broad-band spectrum must be successfully fitted in order to estimate the line parameters.

\section{Fast time variability}

\begin{figure}
   \includegraphics[scale=0.40]{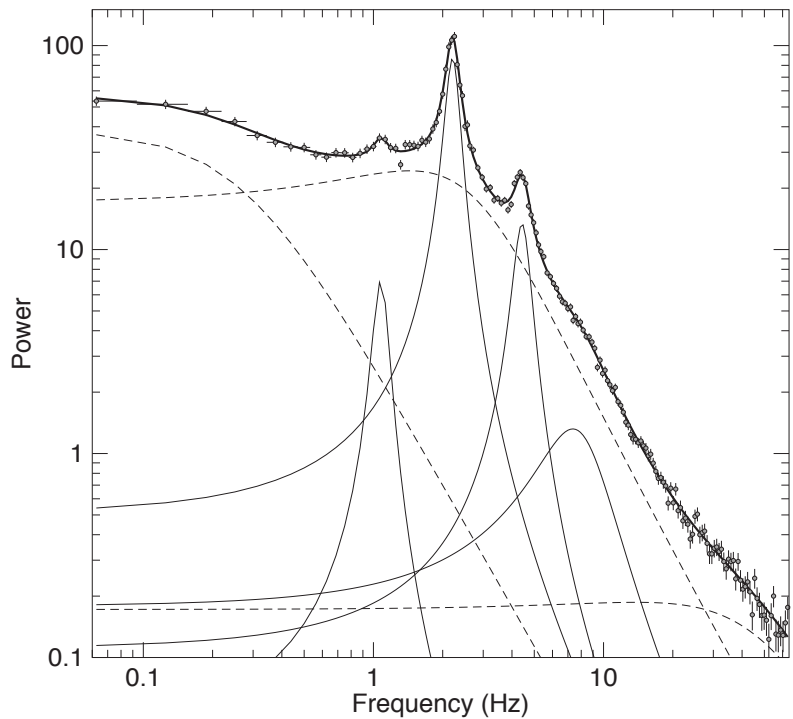}
\caption{PDS from the source GRS 1915+105. Several low-frequency components can be seen. The frequencies of the four peaked ones are in harmonic relation.\cite{ratti}.}
\label{fig:peaks1}
\end{figure}

\begin{figure} 
   \includegraphics[scale=0.40]{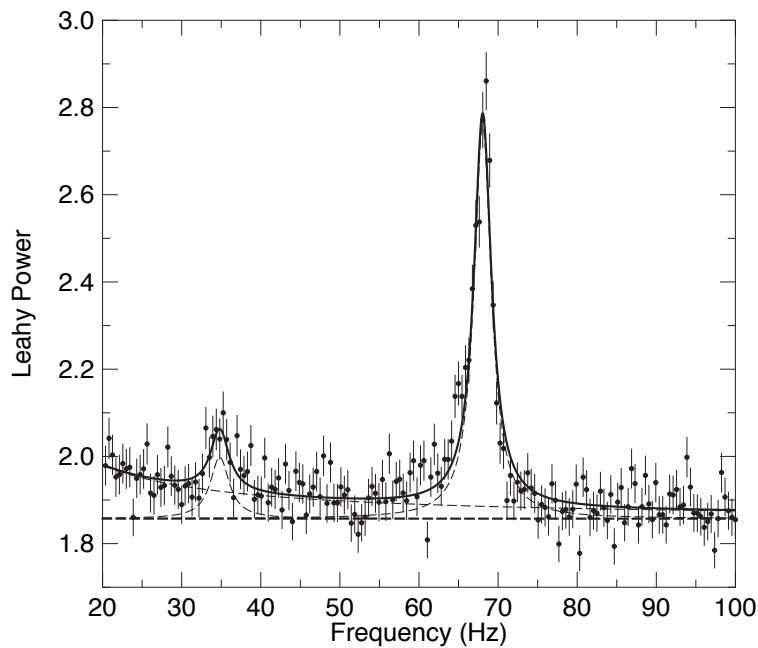}
\caption{Another PDS from GRS 1915+105 in a different state, where high-frequency QPOs are observed\cite{TMBDiego13b}. Two peaks are observed here, with harmonically related frequencies.}
\label{fig:peaks2}
\end{figure}

The X-ray emission observed from BHBs is very variable. Throughout the evolution of an outburst of a transient source, when accretion rate increases and X-ray luminosity reaches values of $10^{36-39}$erg/s, different states are identified, which correspond to very different energy spectra and fast time variability. One of these states, the already mentioned high-soft state, when the emitted spectrum is almost a pure thermal component (that is used by the continuum methods), shows very little fast variability, but the others can see variability up to $\sim$10\% fractional variability. This variability is in the form of broad-band noise and of peaked components called Quasi-Periodic Oscillations (QPO). These peaks are broader than coherent oscillations, but yield very precise measurements of characteristic time scales. In Figs. \ref{fig:peaks1} and \ref{fig:peaks2} one can see two examples. In Fig. \ref{fig:peaks1} there are four peaks in addition to noise, but they are harmonically related so only one frequency (that of the strongest peak) is usually considered. In Fig. \ref{fig:peaks2}, two peaks at higher frequency are seen, with low noise. They are also harmonically related.

It was clear from the first attempts of theoretical modelling of X-ray emission from X-ray binaries that variability could be produced by inhomogeneities in the inner accretion flow. If inhomogeneities can live longer than the orbital time scale at their radius, this could lead to a signal concentrated around the frequencies corresponding to that radius. In this way the inhomogeneity could be used as a ``test particle'' to probe the accretion flow and at the same time the space time close to a black hole. For instance observing high frequencies, the position (and existence) of the ISCO could be measured. 
After these ideas were put forward, QPOs were observed, first in neutron-star binaries then in BHBs. These provided precise frequencies to input to models, although the situation is complicated by the fact that an accretion flow is not made of tennis balls, but is a complex structure of orbiting plasma with its own characteristic time scales, which in principle could also lead to observable signals. 

In the 1990s, multiple QPO peaks started being observed, thanks to NASA's RossiXTE satellite for X-ray astronomy, designed for analysis of fast variability\cite{vdk2006}. Dealing with more than one time scale involves adding different frequencies in addition to the orbital ones. This led to the development of models based on fundamental frequencies of matter orbiting a collapsed object: in addition to the Keplerian one ($\nu_\phi$), the radial epicyclic frequency ($\nu_r$) and the vertical epicyclic frequency ($\nu_\theta$)were considered. For a test particle orbiting at radius $r$ on a slightly eccentric orbit slightly tilted from the plane perpendicular to the spin of a Kerr black hole with specific angular momentum $J$, these are:
$$
\begin{aligned}
   \nu_\phi &= \sqrt{GM/r^3} / 2\pi\, (1+a(r_g/r)^{3/2}) \nonumber \\
   \nu_r^2 &= \nu_\phi^2\,[1-6(r_g/r)+8a\, (r_g/r)^{3/2} - 3a^2\,(r_g/r)^2] \nonumber \\
   \nu_\theta^2 &= \nu_\phi^2\, [1-4a\, (r_g/r)^{3/2} + 3a^2\, (r_g/r)^2]
\end{aligned}
$$

where $r_g=GM/c^2$ and $a=Jc/GM^2$. In addition, possible candidates are also the periastron precession frequency $\nu_{per}=\nu_\phi-\nu_r$ and the nodal precession frequency $\nu_{nod} = \nu_\phi-\nu_\theta$.
In Newtonian approximation $\nu_\phi$, $\nu_r$ and $\nu_\theta$ are identical, but in a strong gravitationa field they are not: $\nu_\phi$ and $\nu_\theta$ decrease with radius, but $\nu_r$ is null at ISCO, has a maximum at a specific radius, then decreases at larger radii. Clearly, matching the observed frequencies with the values from these equation offer the possibility of testing important predictions of GR in the strong-field regime. Notice that the equations above depend only on three parameters: the mass and spin of the black hole and the radius of the orbit. A measurement of three frequencies, if they can be associated to these physical quantities, would give a direct measurement of mass and spin.

\section{Observations}

Going into the intricacy of source states would be beyond the scope of this chapter. I will present only the basic information that can be used to apply theoretical models. Most of the variability that is observed takes place at low frequencies, which we can define as those below $\sim10-20$ Hz, although the most interesting signals are those observed at higher frequencies, in the range where $\nu_\phi$ and $\nu_r$ are expected for a stellar-mass black-hole.

\subsection{Low frequencies}

\begin{figure}
   \includegraphics[scale=0.50]{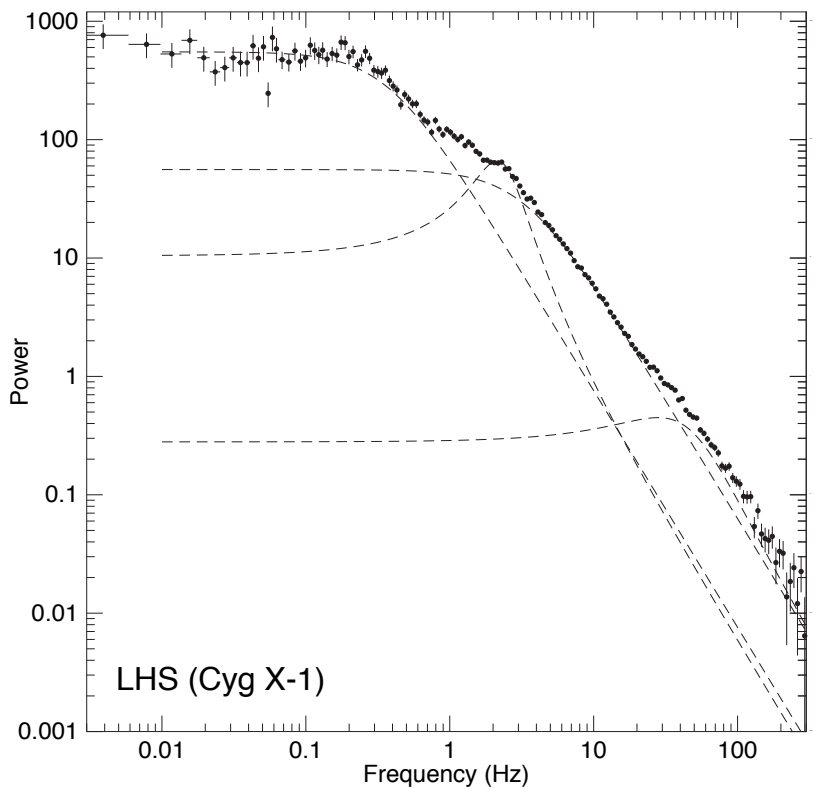}
\caption{PDS from the hard state of Cyg X-1. The dashed lines mark the different noise components.
}
\label{fig:pds1a}
\end{figure}

When BHBs are in their hard state (where the thermal disk described above is not observed and the emission is dominated by other components), strong variability is observed in the form of noise, although sometimes a more peaked component, a QPO, can also be observed. The typical Power Density Spectrum is that shown in Fig. \ref{fig:pds1a}. Four components are seen here, one of which slightly peaked. By applying a Lorentzian model for these components, it is possible to extract a characteristic frequency from each of them\cite{BPK}.

\begin{figure}
   \includegraphics[scale=0.58]{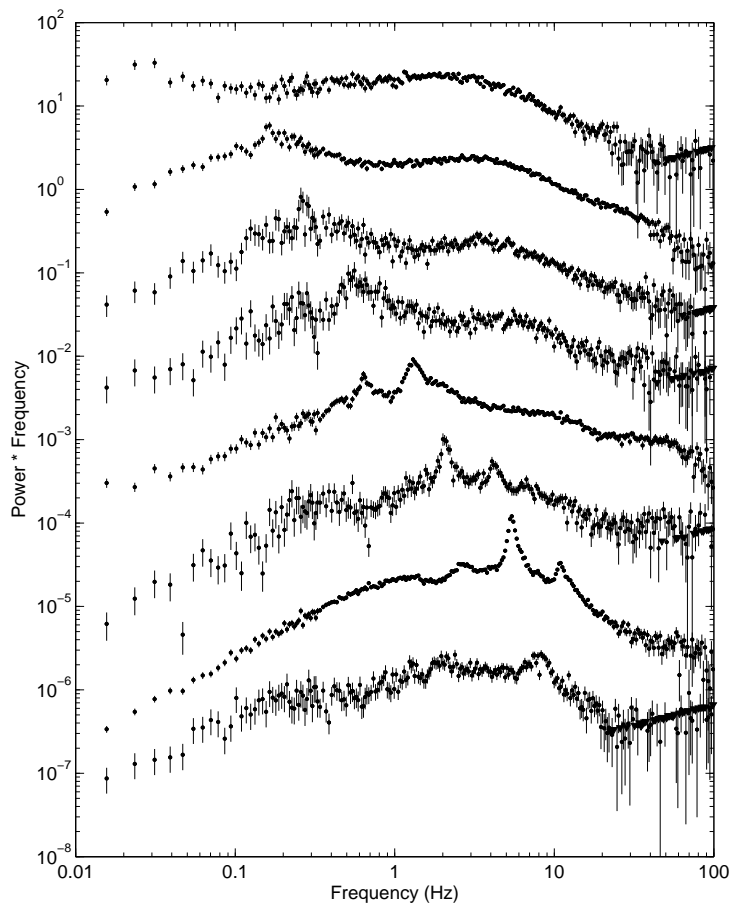} 
\caption{Time sequence (from top to bottom) of PDS (multiplied by frequency) of GX 339-4. The evolution from noise to noise+QPO and the increasing frequencies is evident. From Belloni et al. (2006)\cite{belloni2005}.
}
\label{fig:pds1b}
\end{figure}

As the X-ray flux varies, these frequencies change in a correlated way. Recently, the lowest-frequency component has been associated with propagation effects in the accretion flow, where variability from all radii in the disk contributes to the noise\cite{arevalo,Ingram15}, although the frequency of this component is correlated with the others, an indication that the situation could be more complicated.

When the energy spectrum softens and a disk component starts appearing in the 2-10 keV band, the PSD is consistent with being a high-frequency extension of the one in Fig. \ref{fig:pds1a}, meaning that all frequencies are higher and the total fractional variability is lower. However, a clear peaked component appears, called type-C QPO (see Fig. \ref{fig:peaks1} and Fig. \ref{fig:pds1b}). As the name suggests, there are at least two other types of QPOs (type-A and -B), but the important one to consider in this context is type-C, the most common one. These are QPO peaks whose centroid frequency varies in the 0.01-30 Hz range. They are rather narrow, with a FWHM roughly a tenth of their centroid frequency and can be as strong as 5-10\% fractional rms. They appear often with harmonically related peaks (see Fig. \ref{fig:peaks1}) and their centroid correlates with the energy spectrum, being highest when the spectrum is softest. Their energy spectrum is too hard for the modulated emission to originate from the thermal accretion disk, but the large changes in frequency suggest that the associated radius of emission changes. A strong correlation between their centroid frequency and the characteristic frequency of the lowest noise component in the noise has been discovered\cite{WK}.
Type-C QPOs have been observed from many sources and appear to be very common. Their frequency is too low to be associated to Keplerian orbits, but it can be linked to precession.
In softer states either other types of QPO are observed or, in the full-fledged high-soft state, there is little variability.

\subsection{High frequencies}

While signals at low frequencies are almost always  observed, high-frequency features appear to be rare (or weak). Only with the RossiXTE mission we have started to sample efficiently this frequency range, which led to the discovery of the (related) kHz QPO in neutron-star binaries, of which there are many detections. For black holes, the situation is not so good. The first observation of a high-frequency QPO from a BHB was at 67 Hz from the very variable source GRS 1915+105\cite{Morgan}. After that detection, in sixteen years if operation RXTE discovered very few more from other sources, although GRS 1915+105 showed many (see below)\cite{belloni2012,TMBDiego13a}. Some reported detections are at low statistical significance and others were too broad to be classified as QPOs. We are left with {\it six} sources with at least a significant detection, see Tab.\ref{tab:1}. All these correspond to observations at very high luminosity and therefore strong signal; it is unclear whether this is an effect of QPOs being stronger at high luminosity or our sensitivity being too low at lower luminosities.

\begin{table}
\caption{High-frequency QPOs from BHBs.}
\label{tab:1}       % Give a unique label
\begin{tabular}{lccc}
\hline\noalign{\smallskip}
Source & N$_{peaks}$ & Simultaneity & Frequency (Hz)  \\
\noalign{\smallskip}\hline\noalign{\smallskip}
GRO J1655-40   & 2 & Y & $\sim$300, $\sim$400 \\
XTE J1550-564  & 2 & N & $\sim$180, $\sim$280 \\
XTE J1650-500  & 1 & -  & $\sim$250 \\
H 1743-322        & 2 & N & $\sim$160, $\sim$240 \\
IGR 17091-3624& 1 & -   & $\sim$66 \\
GRS 1915+105  & 4  & Y & $\sim$27, $\sim$34, $\sim$41, $\sim$67 \\
\noalign{\smallskip}\hline
\end{tabular}
\end{table}

Obviously these peaks are not as easy to detect as type-C QPOs, but in addition they seem to be incompatible with the presence of type-C QPOs. They only appear in different states from that of type-C QPOs and there are only a few cases of simultaneous detection (see below). As can be seen from Tab. \ref{tab:1}, three sources (I will discuss the special case of GRS 1915+105 separately) have shown two peaks, although only in one case there are two significant detections in the same observation. For these three systems, the frequencies of the two peaks are close to being in a 3:2 ratio. For GRS 1915+105 things are, as usual, more complicated and no 3:2 ratio is present among the detection in Tab. \ref{tab:1}, although other small integer ratios can be extracted.

The case of GRS 1915+105 is different. This system is very peculiar and, unlike normal transient systems, has been very bright throughout the whole RossiXTE period, which means that it has been observed many times at very high accretion. Indeed, a systematic search through the archive has lead to 51 detections of high-frequency QPOs, out of which 48 have a centroid frequency in the 63-71 Hz range\cite{TMBDiego13a}. This frequency is a very stable number in this system. However, the most promising candidate for its interpretation, namely $\nu_\phi$ @ ISCO is not an option, as the current mass measurements predict a higher frequency even in the case of zero spin.

\section{Models}

The presence of noise in the X-ray emission of X-ray binaries was determined with the first observations from space with the {\it Uhuru} satellite, when relatively long observations could be obtained. The quality of the data did not allow precise determination of the noise properties and early models concentrated on a possible shot-noise nature of the variability\cite{terrell,nolan,BH}, which we now know can be excluded due to statistical properties of the time series. The first QPO peaks in the 1980s were discovered at low frequencies from NS binaries and led to the development of models that involved the spin of the neutron star\cite{lamb}. The discovery of QPOs from BHBs, where these models cannot be applied, changed little. However, with the discovery of high-frequency features with RossiXTE in the second half of the 1990s changed everything and led to the development of models that could be applied to both classes of systems, independent of the nature of the compact object. 

\subsection{The Relativistic Precession Model}

\begin{figure}
   \includegraphics[scale=0.36]{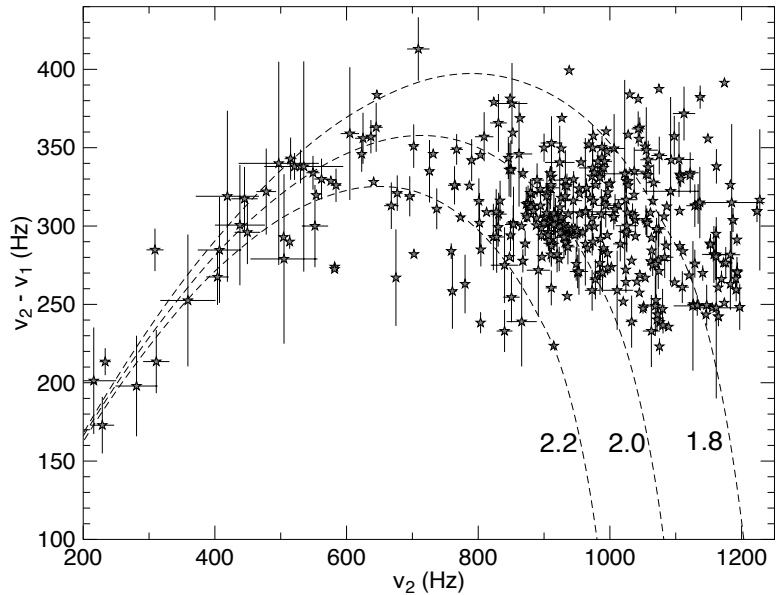}
\caption{Difference between the frequencies of the two kHz QPO peaks in NS binaries as a function of the upper frequency, where all published pairs of QPOs have been included. This plot is a new version of earlier published ones\cite{StellaVietri99,Boutloukos}. The lines correspond to RPM predictions for three values of the neutron-star mass.
}
\label{fig:kHz1}
\end{figure}

The presence of three main features in the variability of neutron-star models and the availability of a large number of detections, led to the search for models that could explain all the QPOs (there are additional types of QPOs and sideband peaks, but it is the three peaks, one at low frequencies and two at high frequencies (called kHz QPOs because of their high frequency that can reach 1000Hz) that can form the base for a successful model. The first such a model is the Relativistic Precession Model (RPM)\cite{StellaVietri98,StellaVietri99,Stella99}. It is a local model, in the sense that the oscillations are associated to a specific radius in the accretion flow, and as most models it only aims at the interpretation of the observed frequencies and does not address a production mechanism. 
In the RPM, the low-frequency QPO is interpreted as $\nu_{nod}$ and the two high-frequency ones as  $\nu_{per}$ and $\nu_\phi$. The frequencies are associated to a single radius around the neutron star. All three GR frequencies outlined above are involved. When applied to the available kHz QPO data, the model did not provide a statistically good fit, but yielded frequencies in the observed range and with the basic dependence between each other. Figure \ref{fig:kHz1} shows the difference between the frequencies of the two kHz QPO peaks (in the RPM corresponding to the radial epicyclic frequency $\nu_r$) as a function of the upper frequency (corresponding to $\nu_\phi$) from all published pairs of kHz QPOs published until now. The lines correspond the model prediction for different neutron star masses (the neutron star rotation has a small effect on these plots), where different points along the line correspond to different radii of emission. The model lines in Fig.\ref{fig:kHz1} go though the cloud of points and the three qualitative predictions of the model are observed: positive correlation at low frequencies (the points on this branch come from the only source where low frequencies were observed, Circinus X-1\cite{Boutloukos}), negative correlation at high frequencies, no $\nu_r$ value above 400 Hz.

\begin{figure}
   \includegraphics[scale=0.36]{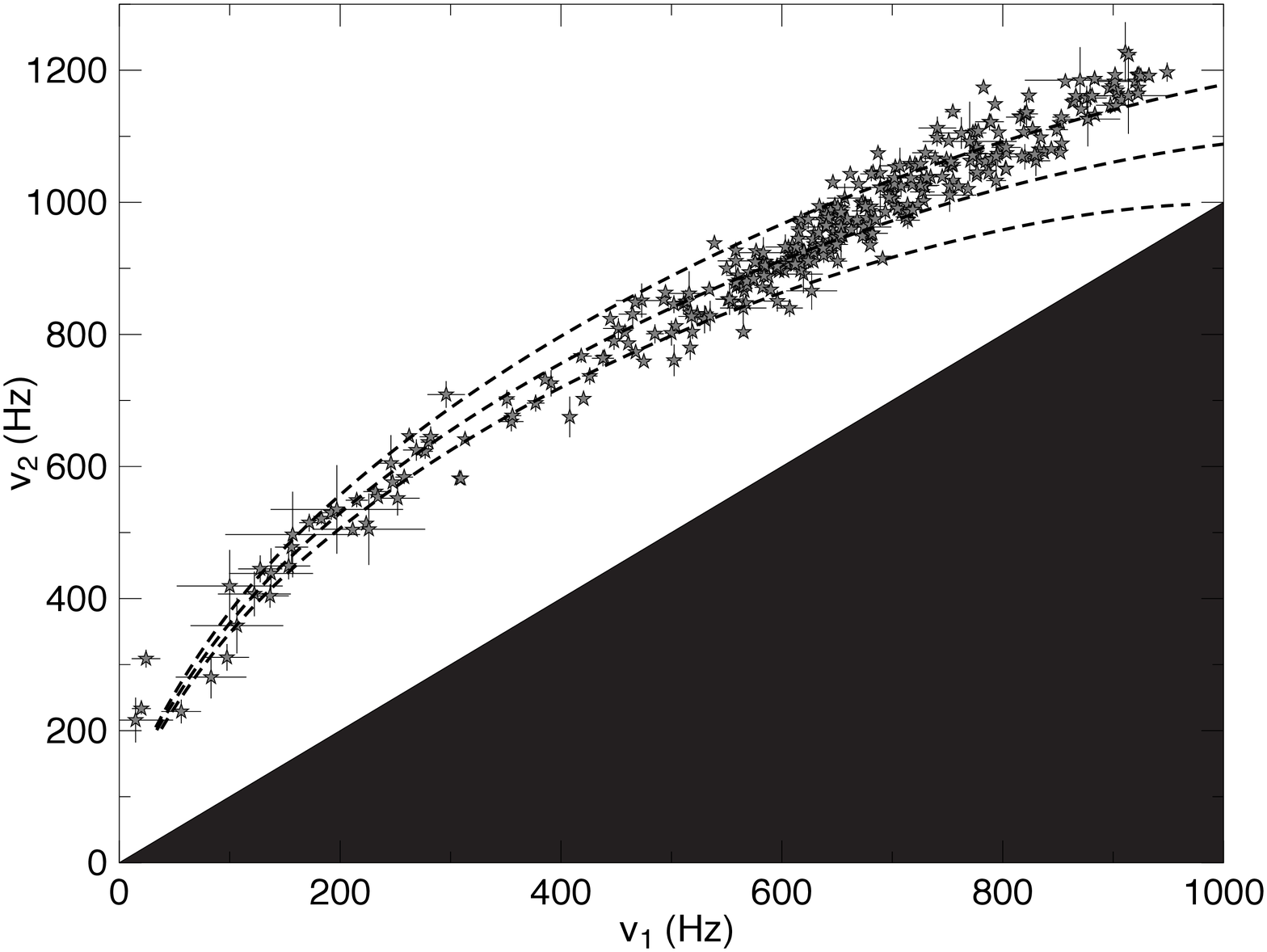}
\caption{Plot of one kHz QPO frequency versus the other. The dark area corresponds to a lower frequency higher than the upper one. The dashed line lines correspond to the same model lines in Fig.\ref{fig:kHz1}.
%From Belloni et al. (2007).\cite{Belloni 2007}
}
\label{fig:kHz2}
\end{figure}

A more direct to look at the data is to plot one kHz QPO frequency versus the other (see Fig.\ref{fig:kHz2}) where different subclasses of NS binaries are shown. There are deviations at high frequencies, but the 2M$_\odot$ model is a viable representation of the data.

In addition to the high-frequency features, the RPM interprets the low-frequency QPO seen in NS binaries and BHBs (also for NS there are more types of QPOs, but one of them can be associated to the one seen from BHBs) as $\nu_{nod}$. In the weak field approximation, this frequency scales with the square of $\nu_\phi$ (the Lense-Thirring effect). This dependence has been observed in NS binaries\cite{StellaVietri98,Psaltis1999}

One QPO in one source stands out: a peak in the range 35-50 Hz has been detected from a NS source that also showed a kHz QPO. These two features are inconsistent with being RPM frequencies originating from the same radius. More observations are needed here to confirm the nature of the low-frequency QPO and the mismatch with the model, but this will have to wait for the source to be observable again, as it was a transient system.

\begin{figure}
   \includegraphics[scale=0.064]{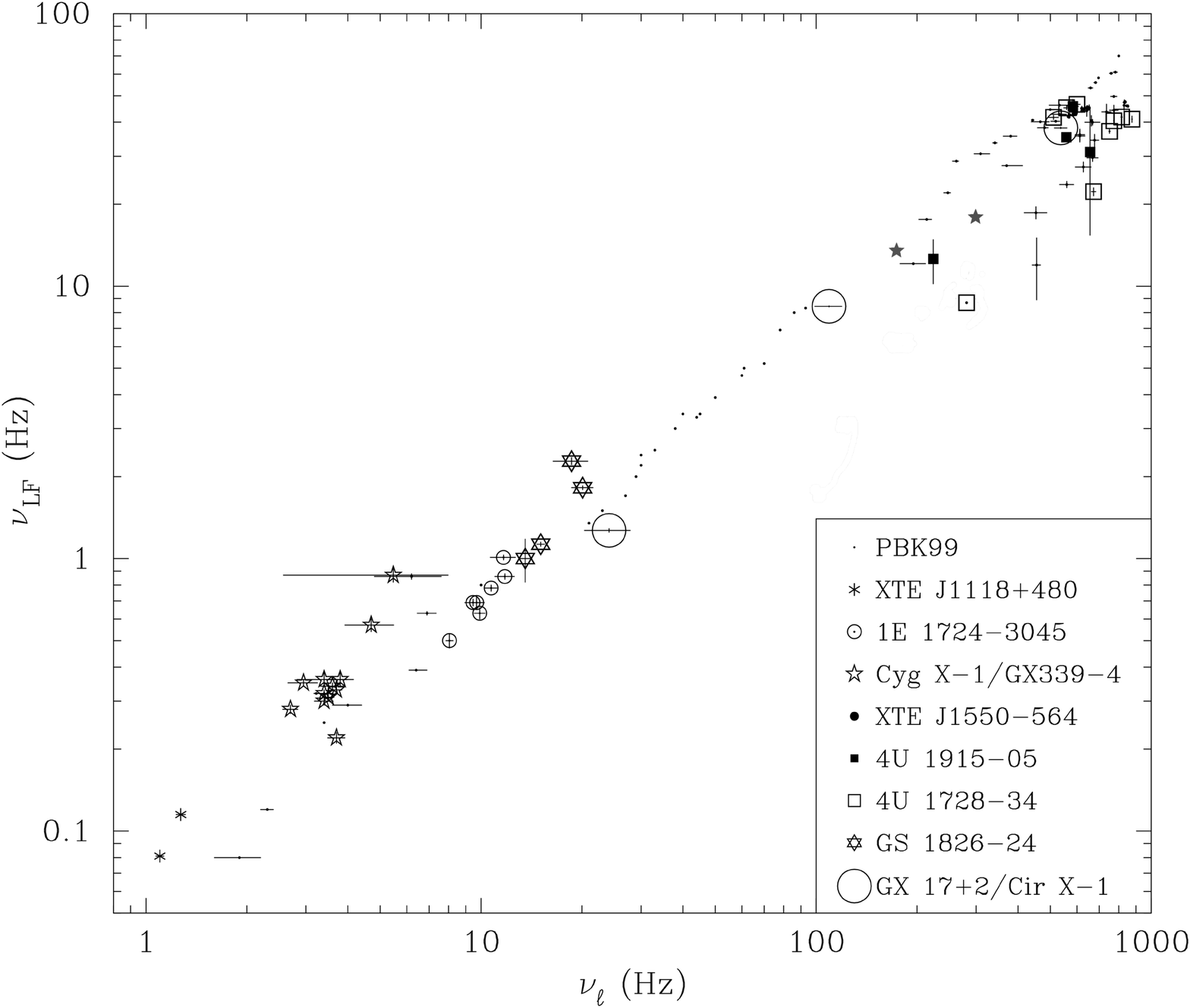} 
\caption{The PBK correlation as published in its second version\cite{BPK}. Some points have been removed as they are not significant and two points have been added, the stars in the upper right of the diagram, corresponding to the detections discussed below.
}
\label{fig:BPK}
\end{figure}

Most of the attention has been on NS binaries, since as shown above the few detections of high-frequency features in BHBs did not correspond to detections of low-frequency QPOs. However, there is an observed correlation that suggests that high-frequency peaks could be observed also at low frequencies (associated to larger radii) and in that case they do not appear as peaks: the so called PBK correlation\cite{PBK,BPK}. The correlation is shown in Fig. \ref{fig:BPK}: this is the original correlation\cite{BPK}, but some incorrect points have been removed and two new points have been added (see below). Here non-homogeneous quantities are plotted. For kHz QPOs, the low-frequency QPO frequency is plotted vs. that of the low-frequency QPO. For sources in the hard states (see PDS in Fig. \ref{fig:pds1a}), both NS and BH, the characteristic frequency of the broad shoulder above it (see Fig. \ref{fig:pds1a}, where it is around 2 Hz) vs. the low-frequency QPO. In other words, while the $x$ axis is always the frequency of the low-frequency QPO, the $y$ axis for $x<20$ Hz represents a broad component, while for $x>100$ Hz it is a high-frequency peak. The points in between are from the same Circinus X-1 source that allowed to measure the left portion of Figs. \ref{fig:kHz1} and \ref{fig:kHz2}.
This correlation appears very tight and connects NS ad BH sources, as well as narrow and broad components, clearly suggesting a common physical mechanism. Indeed the RPM does interpret this correlation without needing additional parameters\cite{Stella99}.

It is interesting to note that this correlation, which covers three orders of magnitude, has been extended by two more orders of magnitude by including frequencies of dwarf-nova oscillations observed in the optical band from cataclysmic variables, binary systems containing white dwarfs\cite{Warner2003}. While the correlation between the WD frequencies is good and it connects to the BPK, it is not clear how to link signals detected in different bands and from different systems, as the X-ray emitting region of an X-ray binary does not exist in a cataclysmic variable, dure to the size of the star.

\subsection{The Epicyclic Resonance Models}

As mentioned above, some of the observed pairs of high-frequency oscillations in BHBs have been observed to be close to a 3:2 ratio. This has led to the development of another class of local models, where the special radius associated to the QPOs is identified by resonances. In particular, there is a radius at which the radial $\nu_r$ and vertical $\nu_\theta$  epicyclic frequencies have simple integer ratios (2:1, 3:2). This can lead to resonance\cite{KA1,KA2,AK}.
Additionally, also a resonance between $\nu_r$ and $\nu_\phi$ has been considered\cite{torok}.
The few available detections of QPO pairs in BHBs do not allow to check further the validity of the model, but the fact that the observed frequencies are observed only to show little variability is naturally explained within these models, since for a given object the resonance radii cannot change. Of course sharp changes from one resonance to the next one can in principle be observed, but no continuous shifts. Indeed, this model cannot be easily applied to neutron stars, where the frequencies vary and span a rather large range, without invoking an additional unknown mechanism that brings these changes, at which point one of the main predictions of the model does not exist anymore. Moreover, these models do not include low-frequency oscillations.

\subsection{Other models}

\begin{figure}
\begin{center}
   \includegraphics[scale=0.15]{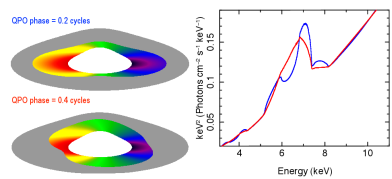} 
   \end{center}
\caption{Left: Ray-tracing representation of the illuminated part of the disk by the precessing inner hot flow at two different QPO phases for observations of H~1743-322. Right: corresponding best fit line profiles\cite{ingramline2}.
}
\label{fig:ingram}
\end{figure}

Other models have been proposed and most of them are based on some of the fundamental frequencies of motion shown above. Both the RPM and the resonance models offer an interpretation for the observed centroid frequencies, but they do not include the presence of an accretion flow, nor they address the nature of the modulated emission. An extended approach to the study of the low-frequency QPOs in the Lense-Thirring hypothesis was presented a few years ago\cite{IngramDoneFragile,ID11,ID12,IM14}. To go beyond the single radius-test-particle approach, the model considers the precession of an extended hot region in the accretion flow, from which the observed QPO frequency would arise from a solid body-like nodal precession that would depend on the  outer radius of the region, located inside a truncated disk. The model allows the interpretation of observed correlations and a modulation of the iron line centroid in H~1743-322 has been observed to match its prediction due to variable irradiation of the accretion disk by the inner region (see Fig. \ref{fig:ingram})\cite{ingramline,ingramline2}.

A model (AEI, Accretion-Ejection Model) has been proposed based on a disk instability involving a spiral instability driven by magnetic stress from a poloidal field. The spiral extracts energy from the disk and transfers it to a corona, which then powers the ejection of relativistic jets\cite{tagger}. The model has also been extended to interpret high-frequency QPOs\cite{tagger2}.

Global disk oscillation models, involving different oscillation modes have been proposed and recently the first instances of QPO signals have been reproduced through hydrodynamical and MHD simulations. The discussion of these models goes beyond the scope of this article and can be found elsewhere\cite{bellonistella}.

\section{Where we stand}

\begin{figure}
   \includegraphics[scale=0.28]{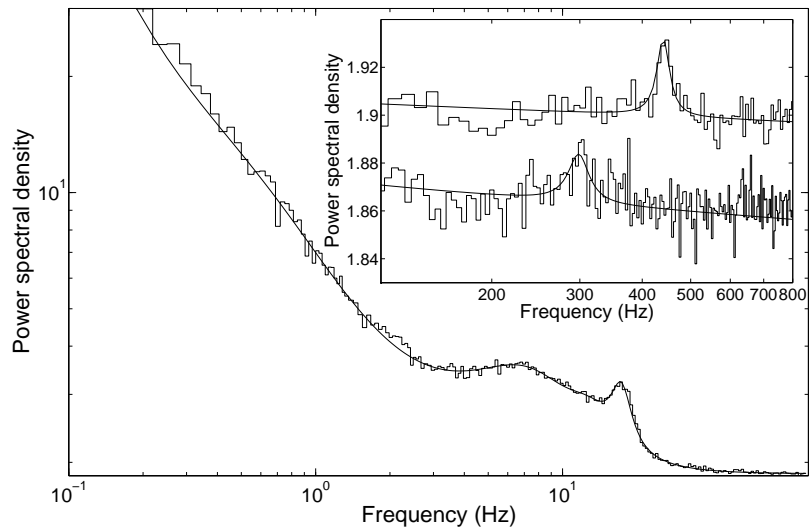}
\caption{Left panel: PDS of the RossiXTE observation of GRO J1655-40 with three simultaneous oscillations. The main panel shows the 18 Hz low-frequency QPO, the inset shows the two high-frequency counterparts (in two different energy bands)\cite{motta1}.
}
\label{fig:3peaks1}
\end{figure}

It is clear that the existing data on high-frequency QPOs is so scarce that it is difficult to test the models above. Low-frequency QPOs have been widely observed, but their tendency not to be detected together with their high-frequency counterparts makes the testing of a full scenario problematic.
However, out of the full archive of RossiXTE there exists one observation of a bright system, GRO J1655-40, where all three QPOs (one at low frequencies and two at high frequencies) have been detected (see Fig.\ref{fig:3peaks1})\cite{motta1}. 

\begin{figure}
   \includegraphics[scale=0.23]{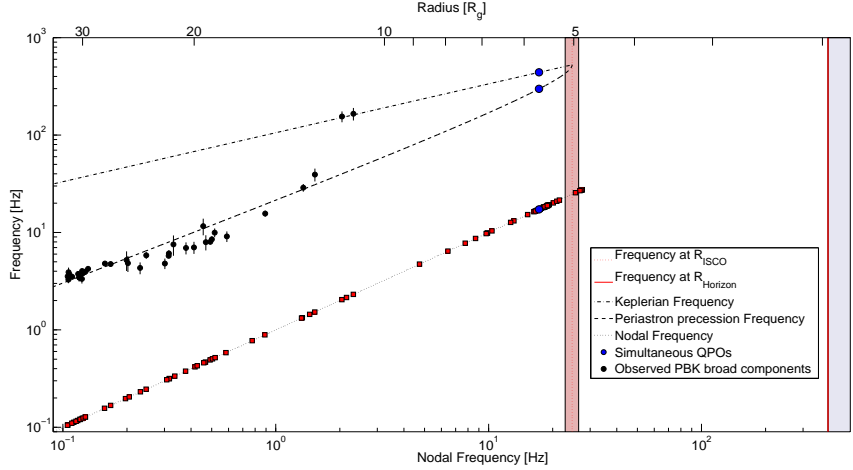}
\caption{Frequency correlation for the RossiXTE observations of GRO J1655-40 with QPOs, with the low-frequency in the abscissa\cite{motta1}. The red points mark low-frequency QPOs (therefore they mark a 1:1 line), the other points represent high-frequency QPOs and broad features (see text).The blue points are the three frequencies visible in Fig,\ref{fig:3peaks1}. The dashed lines are the RPM predictions based on the three blue points, while the red band indicates the range of the expected maximum frequencies for the oscillations due to the presence of the ISCO. The upper x-axis shows the radii corresponding to the frequencies on the lower-axis.
}
\label{fig:3peaks2}
\end{figure}

With the equations above, under the assumption that the three frequencies correspond to GR frequencies at the same radius $R$, it was possible to estimate $R$ and the mass and spin of the black hole with high accuracy: $R=5.677\pm 0.035$ r$_g$, $M=5.307\pm 0.066$ M$_\odot$, $a=0.286\pm 0.003$. Despite the small error bars, the derived mass matches the value obtained from dynamical measurements in the optical, $m=5.4\pm 0.4$ M$_\odot$\cite{motta1,beer}. Since the only two parameters for the black hole are estimated, this also allows to derive a value for the ISCO $r_{ISCO} = 5.031\pm 0.009$ r$_g$.
There are no other instances of three simultaneously detected QPOs, but there are many detections of the low-frequency QPO in GRO J1655-40 spanning the range 0.01-27 Hz. Since spin and mass are fixed, it was possible to test the hypothesis that the broad components observed together with low-frequency QPO represent relativistic frequencies (the PBK correlation, see above). Figure \ref{fig:3peaks2} shows that they fit very well the expected correlation with the low-frequency QPO\cite{motta1}. Notice that although the low-frequency QPO span a very large range, they do not exceed the range of values expected at ISCO, which are not expected to be present.
Moreover, the relative width of the three peaks can be explained with a modest jitter of the radius $R$, making the result even more robust.

\begin{figure}
   \includegraphics[scale=0.23]{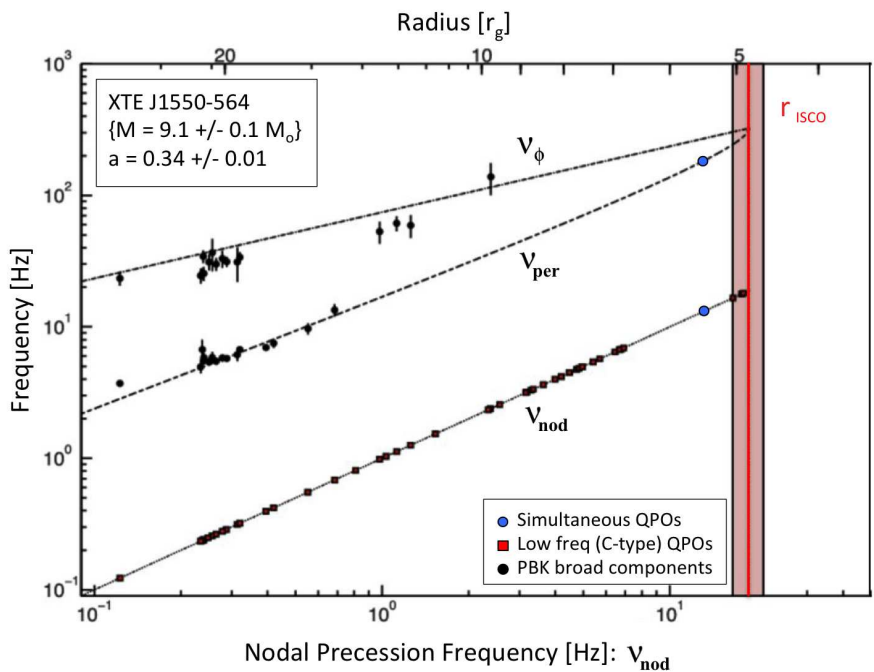}
\caption{Frequency correlation for the RossiXTE observations of GRO J1550-40, same plot as Fig. \ref{fig:3peaks2}\cite{motta2}.
}
\label{fig:3peaks1550}
\end{figure}

As the estimate of the ``correct'' black-hole mass strengthens the association between observed and predicted frequencies, it was possible to apply the model to the second best observation, when another bright system (XTE J1550-564) was observed to show {\it two} simultaneous QPOs, one at low frequencies and one at high frequencies. With two frequencies the equations cannot be solved, but the inclusion of the optically measured black hole mass leads to the estimate of $a=0.34 \pm 0.01$\cite{motta2}. Also in this case, the extension to lower frequencies through broad features is well interpreted by the model (see Fig. \ref{fig:3peaks1550}), as well as the relative width of the two peaks. 

\section{Conclusions}

After years of observations of low- and high-frequency variability from BHBs we now have a number of models to test, but unfortunately too few detections of high-frequency QPOs to reach a firm conclusion. It is very likely that this is only  problem of sensitivity, as also shown by the fact that all detections were obtained when the sources were in their brightest stages. If the results shown in the previous section are confirmed (they are based on two observations only), the RPM model has led to the direct measurement of a stellar black hole mass and therefore to the direct evidence of the presence of a black hole in the system, as well as to the existence of an ISCO and the confirmation of the relativistic precession frequencies in the strong-field regime.

What is needed now is more realistic models and new more sensitive observations. Models that take into account the presence of an accretion flow rather than simple test particles are being developed and have been shown above. What is still missing is a solid determination of the emission process that leads to the production of oscillations that can be as strong as 20\% in rms.
At the same time, new instruments are being developed. At the time of writing, the Indian X-ray satellite AstroSat is operative and is yielding data comparable in quality to those of RossiXTE. No new high-frequency oscillations have been reported yet, but observations are in place. 
For the future, larger missions are being proposed, such as eXTP\cite{extp}, with instruments capable of collecting more photons and reaching a much higher sensitivity for fast variations.

%------------------------------------------------------------------------------------------------------------------------------


\begin{thebibliography}{9}

\bibitem{AK} M.A.~Abramowicz, W.~Klu\' zniak, \emph{A precise determination of black hole spin in GRO J1655-40}, 2001, A\&A, 374, L19.\\

\bibitem{arevalo} P.~Ar\'evalo, P.~Uttley, \emph{Investigating a fluctuating-accretion model for the spectral-timing properties of accreting black hole systems}, 2006, MNRAS, 367, 801.\\

\bibitem{beer} M.E.~Beer, P.~Podsiadlowski, \emph{The quiescent light curve and the evolutionary state of GRO J1655-40}, 2002, MNRAS, 331, 351.\\

\bibitem{TMB18} T.M.~Belloni, \emph{X-ray emission from black-hole and neutron-star binaries}, To appear in HIGH TIME RESOLUTION ASTROPHYSICS - XXVII Canary Islands Winter School of Astrophysics (ed. T.~Shahbaz. J.~Casares Vel\'azquez, T.~Mu\~noz Darias), 2018, arXiv:1803.03641\\

\bibitem{BH} T.~Belloni, G.~Hasinger, \emph{Variability in the noise properties of Cygnus X-1}, 1990, A\&A, 227, L33.\\

\bibitem{BPK} T.~Belloni, D.~Psaltis, M.~van der Klis, \emph{A unified description of the timing features of accreting X-ray binaries}, 2002, ApJ, 572, 392.\\

\bibitem{belloni2005}T.~Belloni, J.~Homan, P.~Casella, et al., \emph{The evolution of the timing properties of the black-hole transient GX 339Ð4 during its 2002/2003 outburst}, 2005, A\&A, 440, 207.\\

\bibitem{Belloni2007} T.~Belloni, M.~M\'endez, J.~Homan, \emph{On the kHz QPO frequency correlations in bright neutron-star X-ray binaries}, 2007, MNRAS, 376, 1133.\\

\bibitem{belloni2012} T.M.~Belloni, A.~Sanna, M.~M\'endez, \emph{High-frequency quasi-periodic oscillations in black hole binaries}, 2012, MNRAS, 426, 1701.\\

\bibitem{TMBDiego13a} T.M.~Belloni, D.~Altamirano, \emph{High-frequency quasi-periodic oscillations from GRS 1915+105}, 2013, MNRAS, 432, 10.\\

\bibitem{TMBDiego13b} T.M.~Belloni, D.~Altamirano, \emph{Discovery of a 34 Hz quasi-periodic oscillation in the X-ray emission of GRS 1915+105}, 2013, MNRAS, 432, 19.\\

\bibitem{bellonistella} T.~Belloni, L.~Stella, \emph{Fast Variability from Black-Hole Binaries}, 2014, Spa. Sci. Rev., 183, 43.\\

\bibitem{BelloniMotta16} T.M.~Belloni, S.E.~Motta, \emph{Transient Black Hole Binaries}, 2016, ASSL, 440, 61.\\

\bibitem{Boutloukos}  S.~Boutloukos, M.~van der Klis, D.~Altamirano, et al., \emph{Discovery of twin kHz QPOs in the peculiar X-ray binary Circinus X-1}, 2006, ApJ, 653, 1435.\\

\bibitem{CasaresJonker} J.~Casares, P.G.~Jonker, \emph{Mass Measurements of Stellar and Intermediate-Mass Black Holes}, 2014, Spa. Sci. Rev., 183, 223.\\

\bibitem{Davis} S.W.~Davis, I.~Hubeny, \emph{A grid of relativistic, non-LTE accretion disk models for spectral fitting of black hole binaries}, 2006, ApJSuppl., 164, 530.\\

\bibitem{FabianRoss} A.C.~Fabian R.R.~Ross, \emph{X-ray reflection}, 2010, Spa. Sci. Rev., 157, 167.\\

\bibitem{KA1} W.~Klu\' zniak, M.A.~Abramowicz, \emph{The physics of kHz QPOs -- strong gravity's coupled anharmonic oscillators}, 2001, astro\_ph/0105057.\\

\bibitem{KA2} W.~Klu\' zniak, M.A.~Abramowicz, \emph{Parametric epicyclic resonance in black hole disks: QPOs in micro-quasars}, 2002, astro\_ph/0203314.\\

\bibitem{kramer} M.~Kramer, I.H.~Stairs, R.N.~Manchester, et al., \emph{Tests of General Relativity from timing the double pulsar}, 2006, Science, 314, 97.\\

\bibitem{Ingram15} A.~Ingram, \emph{Modelling aperiodic X-ray variability in black hole binaries as propagating mass accretion rate fluctuations: A short review}, 2016, AN, 337, 385.\\

\bibitem{IngramDoneFragile} A.~Ingram, C.~Done, P.C.~Fragile, \emph{Low-frequency quasi-periodic oscillations spectra and Lense-Thirring precession}, 2009, MNRAS, 397, L101.\\

\bibitem{ID11} A.~Ingram, C.~Done, \emph{A physical model for the continuum variability and quasi-periodic oscillation in accreting black holes}, 2011, MNRAS, 415, 2323.\\

\bibitem{ID12} A.~Ingram, C.~Done, \emph{Modelling variability in black hole binaries: linking simulations to observations}, 2012, MNRAS, 419, 2369.\\

\bibitem{IM14} A.~Ingram, S.~Motta, \emph{Solutions to the relativistic precession model}, 2014, MNRAS, 444, 2065.\\

\bibitem{ingramline} A.~Ingram, M.~van der Klis, M.~Middleton, et al., \emph{A quasi-periodic modulation of the iron line centroid energy in the black hole binary H1743-322}, 2016, MNRAS, 464, 2979.\\

\bibitem{ingramline2} A.~Ingram, M.~van der Klis, M.~Middleton, et al., \emph{Tomographic reflection modelling of quasi-periodic oscillations in the black hole binary H 1743-322}, 2017, MNRAS, 461, 1967.\\

\bibitem{lamb}F.K.~Lamb, et al., \emph{Quasi-periodic oscillations in bright galactic-bulge X-ray sources}, 1985, Nature, 317, 681.\\

\bibitem{Jeff} J.E.~McClintock, R.~Narayan, J.F.~Steiner, \emph{Black Hole Spin via Continuum Fitting and the Role of Spin in Powering Transient Jets}, 2014, Spa. Sci. Rev., 183, 295.\\

\bibitem{MillerMiller} M.C.~Miller, J.M.~Miller, \emph{The Masses and Spins of Neutron Stars and Stellar-Mass Black Holes}, 2014, Phys. Rep., 548, 1.\\

\bibitem{Morgan} E.H.~Morgan, R.A~Remillard, J.~Greiner, \emph{RXTE obsrevations of QPOs in the black hole candidate GRS 1915+105}, 1997, ApJ, 482, 993.\\

\bibitem{motta1} S.E.~Motta, T.M.~Belloni, L.~Stella, et al., \emph{Precise mass and spin measurements for a stellar-mass black hole through X-ray timing: the case of GRO J1655-40}, 2014a, MNRAS, 437, 2554.\\

\bibitem{motta2} S.E.~Motta, T.~Mu\~noz-Darias, A.~Sanna, et al., \emph{Black hole spin measurements through the relativistic precession model: XTE J1550-564}, 2014b, MNRAS, 439, L65.\\

\bibitem{nolan} P.L.~Nolan, D.E.~Gruber, J.L.~Matteson, et al., \emph{Rapid variability of 10-140 keV X-rays from Cygnus X-1}, 1981, ApJ, 246, 494.\\

\bibitem{Psaltis1999} D.~Psaltis, R.~Wijnands, J.~Homan, et al., \emph{On the Magnetospheric Beat-Frequency and Lense-Thirring Interpretations of the Horizontal-Branch Oscillation in the Z Sources}, 1999, ApJ, 520, 763.\\

\bibitem{PBK} D.~Psaltis, T.~Belloni, M.~van der Klis, \emph{Correlations in quasi-periodic oscillation and noise frequencies among neutron star and black hole X-ray binaries}, 1999, ApJ, 520, 262.\\

\bibitem{Psaltis} D.~Psaltis, \emph{Probes and Tests of Strong-Field Gravity with Observations in the Electromagnetic Spectrum}, 2008, Living Reviews in Relativity, 11, 9.\\

\bibitem{ratti} E.M.~Ratti, T.M.~Belloni, S.E.~Motta, \emph{On the harmonics of the low-frequency quasi-periodic oscillation in GRS 1915+105}, 2012, MNRAS, 423, 694.\\

\bibitem{RossFabian} R.R.~Ross, A.C.~Fabian, \emph{X-ray reflection in accreting stellar-mass black hole systems}, 2007, MNRAS, 381, 1697.\\

\bibitem{SS}N.I.~Shakura, R.A.~Sunyaev, \emph{Black holes in binary systems. Observational appearance}, 1973, A\&A, 24, 337.\\

\bibitem{Steiner} J.F.~Steiner, R.C.~Reis, J.E.~McClintock, et al., \emph{The spin of the black hole microquasar XTE J1550-564 via the continuum-fitting and Fe-line methods}, 2011, MNRAS, 416, 941.\\

\bibitem{StellaVietri98}L.~Stella, M.~Vietri, \emph{Lense-Thirring precession and qasi-periodic oscillations in Low-Mass X-ray Binaries}, 1998, ApJ, 492, L59.\\

\bibitem{StellaVietri99}L.~Stella, M.~Vietri, \emph{kHz quasiperiodic oscillations in Low-Mass X-ray Binaries as probes of General Relativity in the strong-field regime}, 1999, Phys. Rev. Lett., 82, 17.\\

\bibitem{Stella99}L.~Stella, M.~Vietri, S.M.~Morsink, \emph{Correlations in the quasi-periodic oscillation frequencies of Low-Mass X-ray Binaries and the Relativistic Precession Model}, 1999. ApJ, 524, L63.\\

\bibitem{tagger} M.~Tagger, P.~Varni\`ere, J.~Rodriguez, et al., \emph{Magnetic floods: Magnetic Floods: A Scenario for the Variability of the Microquasar GRS 1915+105}, 2004, ApJ, 607, 410.\\

\bibitem{tagger2} M.~Tagger, P.~Varni\`ere, \emph{Accretion-Ejection Instability, MHD Rossby Wave Instability, Diskoseismology, and the High-Frequency QPOs of Microquasars}, 2006, ApJ, 652, 1457.\\

\bibitem{terrell} N.J.~Terrell, \emph{Shot-Noise Character of Cygnus X-1 Pulsations}, 1972, ApJ, 174, L35.\\

\bibitem{torok} G.~T\"or\"ok, M.A.~Abramowicz, W.~Klu\' zniak, et al., \emph{The orbital resonance model for twin peak kHz quasi periodic oscillations in microquasars}, 2005, A\&A, 436, 1.\\

\bibitem{vdk2006}M.~van der Klis, \emph{Rapid X-ray variability}, \textit{Compact stellar X-ray sources} (W.Lewin \& M. van der Klis eds.), Cambridge Astrophysics Series 39, Cambridge University Press, Cambridge, 2006), p. 39.\\

\bibitem{Warner2003} B.~Warner, P.A.~Woudt, M.L.~Pretorius, \emph{Dwarf nova oscillations and quasi-periodic oscillations in cataclysmic variables - III. A new kind of dwarf nova oscillation, and further examples of the similarities to X-ray binaries}, 2003, MNRAS, 344, 1193.\\

\bibitem{weisberg}J.M.~Weisberg, Y.~Huang, \emph{Relativistic measurements from timing the binary pulsar PSR B1913+16}, 2016, ApJ, 829:55.\\

\bibitem{WK} R.~Wijnands, M.~van der Klis, \emph{The Broadband Power Spectra of X-Ray Binaries}, 1999, ApJ, 522, 965.\\

\bibitem{extp}S.N.~Zhang, et al., \emph{eXTP: Enhanced X-ray Timing and Polarization mission}, 2017, SPIE Proc. 9905.\\

\end{thebibliography}
\end{document}